\documentclass[apj,iop]{emulateapj}
\slugcomment{The Astrophysical Journal, 764:93 (13pp), 2013 February 10} 
\usepackage[table]{xcolor}
\usepackage[normalem]{ulem}
\usepackage{rotating}
\usepackage{tabularx,booktabs}

\begin{document}

\title{Evidence for quasi-periodic X-ray dips from an Ultraluminous X-ray source: \\
Implications for the binary motion}
\author{Dheeraj R. Pasham\altaffilmark{1,2}, Tod E. Storhmayer\altaffilmark{2}}
\affil{$^{1}$Astronomy Department, University of Maryland, College Park, MD 20742; email: dheeraj@astro.umd.edu; richard@astro.umd.edu \\ 
$^{2}$Astrophysics Science Division, NASA's GSFC, Greenbelt, MD 20771; email: tod.strohmayer@nasa.gov \\
{\it Received 2012 March 19; accepted 2012 December 7; published 2013 January 29}
}

\begin{abstract}
We report results from long-term ($\approx$ 1240 days) X-ray (0.3-8.0
keV) monitoring of the ultraluminous X-ray source NGC 5408 X-1 with
the {\it Swift}/X-Ray Telescope. Here we expand on earlier work by
Strohmayer (2009) who used only a part of the present data set.  Our
primary results are: (1) the discovery of sharp, quasi-periodic,
energy-independent dips in the X-ray intensity that recur on average
every 243 days, (2) the detection of an energy-dependent (variability
amplitude decreases with increasing energy), quasi-sinusoidal X-ray
modulation with a period of 112.6$\pm$4 days the amplitude of which
weakens during the second half of the light curve and (3) spectral
evidence for an increase in photoelectric absorption during the last
continuous segment of the data. We interpret the X-ray modulations
within the context of binary motion in analogy to that seen in
high-inclination accreting X-ray binaries. If correct, this implies
that NGC 5408 X-1 is in a binary with an orbital period of 243$\pm$23
days, in contrast to the 115.5 day quasi-sinusoidal period previously
reported by Strohmayer (2009). We discuss the overall X-ray modulation
within the framework of accretion via Roche lobe overflow of the donor
star.  In addition, if the X-ray modulation is caused by vertically
structured obscuring material in the accretion disk, this would imply
a high value for the inclination of the orbit. A comparison with
estimates from accreting X-ray binaries suggests an inclination $\ga$
70$^{\circ}$. We note that, in principle, a precessing accretion disk
could also produce the observed X-ray modulations.
\end{abstract}



\section{Introduction \& Background}

Ultraluminous X-ray sources (ULXs) are bright, point-like,
extragalactic sources with X-ray luminosities in the range from a few
$\times$ 10$^{39}$ ergs s$^{-1}$ to as high as 10$^{41}$ ergs s$^{-1}$
(e.g., Swartz et al. 2004 and references therein). The nature of the
physical process producing such high X-ray output is not completely
understood, but there is now strong evidence that some ULXs are
accretion-powered systems containing a black hole. What remains more
controversial is the mass of the accretor.

Current arguments suggest that these sources are either stellar-mass
black holes (mass range: 3-50 M$_{\odot}$) accreting via a
super-Eddington mechanism (e.g., K\"ording et al. 2002; King et
al. 2001; Begelman 2002), or that they comprise an intermediate-mass
black hole (mass range: a few$\times$(100-1000)M$_{\odot}$) accreting
at sub-Eddington accretion rates (Colbert \& Mushotzky 1999). It is
possible that the population of ULXs is an inhomogeneous sample with
both types of sources present. A search for intermediate-mass black
holes is important as they are often required as building blocks to
explain the formation of the super-massive black holes (Volonteri et
al. 2003; Davies et al. 2011) that reside at the centers of almost all
massive galaxies (Magorrian et al. 1998). In this work we focus on the
ULX NGC 5408 X-1, one of the promising candidates for an
intermediate-mass black hole (Strohmayer \& Mushotzky 2009). Here we
present evidence for the detection of quasi-periodic dips in its X-ray
light curve that likely trace the orbital motion of the system.

There are now two detections of periodicities from long term ($\sim$ a
few hundred days) X-ray monitoring of ULXs which may reflect the
orbital motion of these systems: a 62 day modulation in M82 X-1
detected with RXTE (Kaaret et al. 2007), and a 115.5 day period in NGC
5408 X-1 obtained from {\it Swift} data (Strohmayer 2009, S09
hereafter).  The main results of S09 are: (a) the discovery of a
quasi-sinusoidal modulation of the X-ray flux with a period of
115.5$\pm$4 days and (b) that the modulation amplitude decreases with
increasing energy. The present work utilizes additional data acquired
by {\it Swift} over a much longer temporal baseline (more than twice
as long as used by S09) and can be regarded as an extension of the
work by S09.

The phenomenon of periodic orbital X-ray modulations within the
context of galactic X-ray binaries has been well-studied for over 30
years now (e.g., White \& Holt 1982; Mason 1986). The basic idea is
that there is a distribution of obscuring material around the X-ray
emitting region and as the X-ray source orbits the center of mass of
the binary, our line of sight intercepts varying amounts of the
intervening material resulting in the observed modulation (e.g.,
Parmar \& White 1988; Armitage \& Livio 1998). In addition, these
X-ray variations are expected to recur with the orbital period of the
binary. However, due to the turbulent nature of the accretion process
and irregularities within the material surrounding the X-ray source,
these modulations may not be strictly periodic (see, for example,
Smale et al. 1988; Barnard et al. 2001; Kuulkers et
al. 2012). Nevertheless, they provide an excellent means to track the
orbital motion of the X-ray source.

Moreover, the nature of the obscuring medium, i.e., its distribution
around the X-ray source as well as its density and ionization state,
dictates the observed modulation profiles. In the case of the
high-inclination ($\ga$ 60$^{\circ}$) low-mass X-ray binaries (LMXBs),
periodic decreases in the X-ray flux (X-ray dips) extending over
10-30\% of their orbital phase have been observed (e.g., White et
al. 1995). In these cases it is generally accepted that the X-ray
variations are due to absorption in the ``bulge'' at the edge of the
accretion disk, where the accretion stream from the Roche-lobe filling
companion star impacts the accretion disk (White \& Holt 1982;
Bisikalo et al. 2005). Such variations have been predominantly
observed in the X-ray light curves of high-inclination neutron star
LMXBs, viz., XB 1916-053 (e.g., Boirin et al. 2004), XB 1254-690
(e.g., Smale et al. 2002), EXO 0748-676 (Church et al. 1998).

On the other hand, sharp drops in the X-ray flux lasting less than one
percent of the orbital phase have been observed in a small sample of
accreting black hole binaries. These include the LMXB GRO J1655-40
(Kuulkers et al. 1998, 2000) and the high-mass X-ray binary (HMXB)
Cygnus X-1 (Baluci{\'n}ska-Church et al. 2000; Feng \& Cui 2002). In
the case of GRO J1655-40 the sharp dips last a very small fraction of
the orbital period (a few minutes compared to its orbital period of
2.62 days) and are confined between the orbital phases of 0.7 and 0.9
(phase 0 corresponds to the superior conjunction of the X-ray source,
i.e., when the companion star is in front of the black hole with
respect to our line of sight). The short duration of these dips
suggests that the absorbing medium is likely filamentary in
nature. Based on 3-dimensional numerical simulations of the accretion
stream impacting the disk in compact binaries (Armitage \& Livio
1998), Kuulkers et al. (2000) suggested that the likely scenario
operating in GRO J1655-40 is that the accretion stream from the Roche
lobe overflow of the stellar companion splashes onto the disk rim,
creating a local distribution of material above and below the plane of
the disk. Also, the numerical work by Armitage \& Livio (1998)
predicts that the sharp dips should occur preferentially around the
orbital phase of 0.8 (the stream-disk impact site), similar to those
observed in GRO J1655-40 (Kuulkers et al. 2000). This suggested that
the absorber in the case of GRO J1655-40 is a local distribution
(between orbital phases of 0.7-0.9) of clumps above and below the
accretion disk and is then somewhat different compared to the dipping
in neutron star LMXBs.

The HMXB Cygnus X-1 exhibits two types of ``sharp'' X-ray dips (type-A
and type-B as classified by Feng \& Cui 2002). The type-A dips are
energy-dependent, i.e., accompanied by an increase in the hardness
ratio while the type-B dips are energy-independent, i.e., no evidence
for an increase in the hardness ratio. While the type-A dips are
preferentially distributed roughly about the superior conjunction of
the X-ray source, type-B dips occur randomly over the binary
orbit. The type-A dips are attributed to being produced due to
absorption by density enhancements (clumps or ``blobs,'' in the
vernacular) in an inhomogeneous wind from the companion star. On the
other hand, type-B dips are suggested to be caused by partial covering
of an extended X-ray source by an opaque screen (Feng \& Cui
2002). Numerical simulations of the accretion flow of wind-fed systems
including HMXBs have shown that the tidal force from the compact
object can distort the companion star and give rise to a focused wind
in the direction of the compact source (Blondin et al. 1991). This
tidal wind then develops into a Roche lobe overflow as the surface of
the stellar companion reaches its equipotential surface. The Coriolis
force deflects the accretion stream such that it does not directly
impact the compact companion but goes around it (see Figure 7 of
Blondin et al. 1991). The density in these tidal streams can be as
high as 20-30 times the ambient density and therefore, in principle,
can serve as an opaque absorber of the X-rays (see also,
Baluci{\'n}ska-Church et al. 2000). Furthermore, these simulations
predict that systems with dominant tidal streams should show evidence
for pronounced dipping around the orbital phase of $\sim$ 0.6. A study
of the distribution of the sharp X-ray dips from Cygnus X-1 as a
function of orbital phase has shown that there are indeed two peaks:
the primary peak corresponds to phase 0, i.e., superior conjunction of
the X-ray source and a secondary peak at a phase of 0.6
(Baluci{\'n}ska-Church et al. 2000).

In summary, two types (based on the duration of the individual dips)
of periodic/quasi-periodic X-ray modulations have been observed in
accreting X-ray binaries: 1) broad dips lasting 10-30\% of the orbital
period have been seen in a sample of neutron star LMXBs and 2) sharp
dips lasting less than one percent of the orbital period have been
observed from the black hole LMXB and HMXB, GRO J1655-40 and Cygnus
X-1, respectively. It is important to note that the location of the
absorber is different in the neutron star LMXBs (bulge at the edge of
the accretion disk) compared with GRO J1655-40 (clumps above the
accretion disk) and Cygnus X-1 (accretion stream).

Here we report on a study of the X-ray monitoring data from NGC 5408
X-1 obtained with the {\it Swift} X-Ray Telescope (XRT).  In
particular, we present evidence for sharp dips in the X-ray light
curve and interpret the observed variability in the context of binary
motion. The paper is arranged as follows. In Section 2 we discuss the
details of the {\it Swift}/XRT data used for this work. In Section 3
we study the long-term ($\sim$ a few hundred days) timing behavior of
the source. In particular, we report the detection of the
quasi-periodic, sharp X-ray dips along with the smooth,
quasi-sinusoidal X-ray modulation. Both modulations are consistent
with the dipping phenomena seen in accreting galactic X-ray
binaries. However, to avoid confusion and to be able to clearly
distinguish between the two kinds of modulations we will refer to the
sharp X-ray dips as simply dips or sharp dips while we refer to the
smooth component as the smooth quasi-sinusoidal X-ray modulation.  In
Section 4 we present spectral evidence for a change in the physical
properties of the system. We also study the spectral differences
between the sharp dips and the other portions of the data. We discuss
the implications of our results on the orbital motion of the X-ray
source in Section 5. We summarize our work in Section 6.

\begin{figure*}

\begin{center}
\includegraphics[width=6.5in, height=2.5in, angle=0]{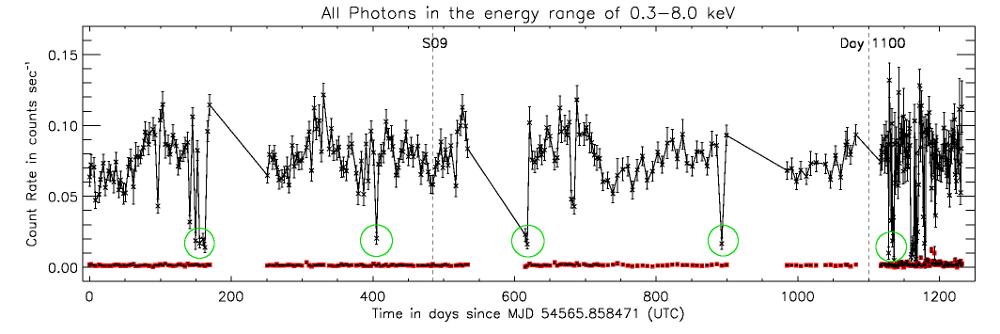}
\end{center}

{\textbf{Figure 1:} The {\it Swift}/XRT X-ray light curve of the ULX
NGC 5408 X-1 in the 0.3-8.0 keV energy band. The background count
rates with their respective error bars are also shown (red points and
black error bars within). The most prominent feature of the light
curve is the quasi-periodic recurrence of the dips. These are
highlighted with green circles. The vertical line labeled as S09 marks
the last observation used by Strohmayer (2009). The vertical line
labeled as ``day 1100'' corresponds to the approximate epoch at which
the physical properties of the system may have changed (see \S 4.1). 
Time zero corresponds to MJD 54565.85847 (UTC).}
\label{fig:figure1}
\end{figure*}
\section{{\it Swift}/XRT Observations}

The X-ray Telescope (XRT) onboard {\it Swift} began monitoring the ULX
NGC 5408 X-1 in 2008 April as part of an approved Cycle 4 program (PI:
Strohmayer).  We include in our analysis all the observations obtained
since the beginning of that program through 2011 August 30. The
observing cadence has varied over this timespan of $\approx$ 3.4 yrs
(1240 days), but on average the source was observed, when viewable,
for a few ks once every 4 days.  Gaps in the coverage due to {\it
Swift} observing constraints occurred from 2008 September 26 - 2008
December 16 (81 days), 2009 September 25 - 2009 December 15 (81 days),
2010 September 26 - 2009 December 20 (85 days) and 2011 March 27 -
2011 May 1 (35 days). This provided a total of 305 pointed
observations with a cumulative exposure of $\approx$ 500 ks
distributed over a temporal baseline of $\approx$ 1240 days.

All the observations were carried out in the photon counting (PC) data
mode. We began our analysis with the level-1 raw XRT event files (data as
stored in the {\it Swift} archive). Each of the event files was
reduced with the standard {\it xrtpipeline} data reduction tool. One
crucial consideration during the reduction process was to mitigate the
impact of bad pixels/columns on the XRT CCD. When a source is
positioned on such pixels, it can lead to an incorrect measurement of
the flux. Furthermore, the bad pixels can result in an erroneous
estimate of the response (effective area) of the instrument. The
University of Leicester's XRT data analysis web page
(http://www.swift.ac.uk/analysis/xrt/exposuremaps.php) provides a
detailed discussion of this problem. The solution is to create
exposure maps that account for the presence of bad pixels. We used
{\it xrtexpomap} ({\it xrtpipeline} with the qualifier {\it xrtexpomap
= yes}) to create exposure maps for each of the individual
observations. These exposure maps were then used to correct the light
curves (using {\it xrtlccorr}) and the ancillary response files
(effective area) (using {\it xrtmkarf}) of each of the observations.

As recommended in the XRT's user guide, we only used events with
grades 0 - 12 for further processing. We then used {\it XSELECT} to
extract light-curves and spectra from the individual observations. We
extracted source light-curves and spectra from a circular region of
radius 47.1'' centered around the source. This particular value was
chosen to include roughly 90\% (at 1.5 keV) of the light from the
source (estimated from the fractional encircled energy of the XRT). A
background region, free of other sources, was extracted in a nearby
region. Given the low individual exposure times, to better estimate
the background, we chose a circular region of twice the radius of the
source, i.e., four times the source area. The same source and
background region was used for all the 305 observations. We present
results from detailed timing and spectral analysis in the following
sections.

\begin{figure*}

\begin{center}
\includegraphics[width=7.0in, height=3.8in, angle=0]{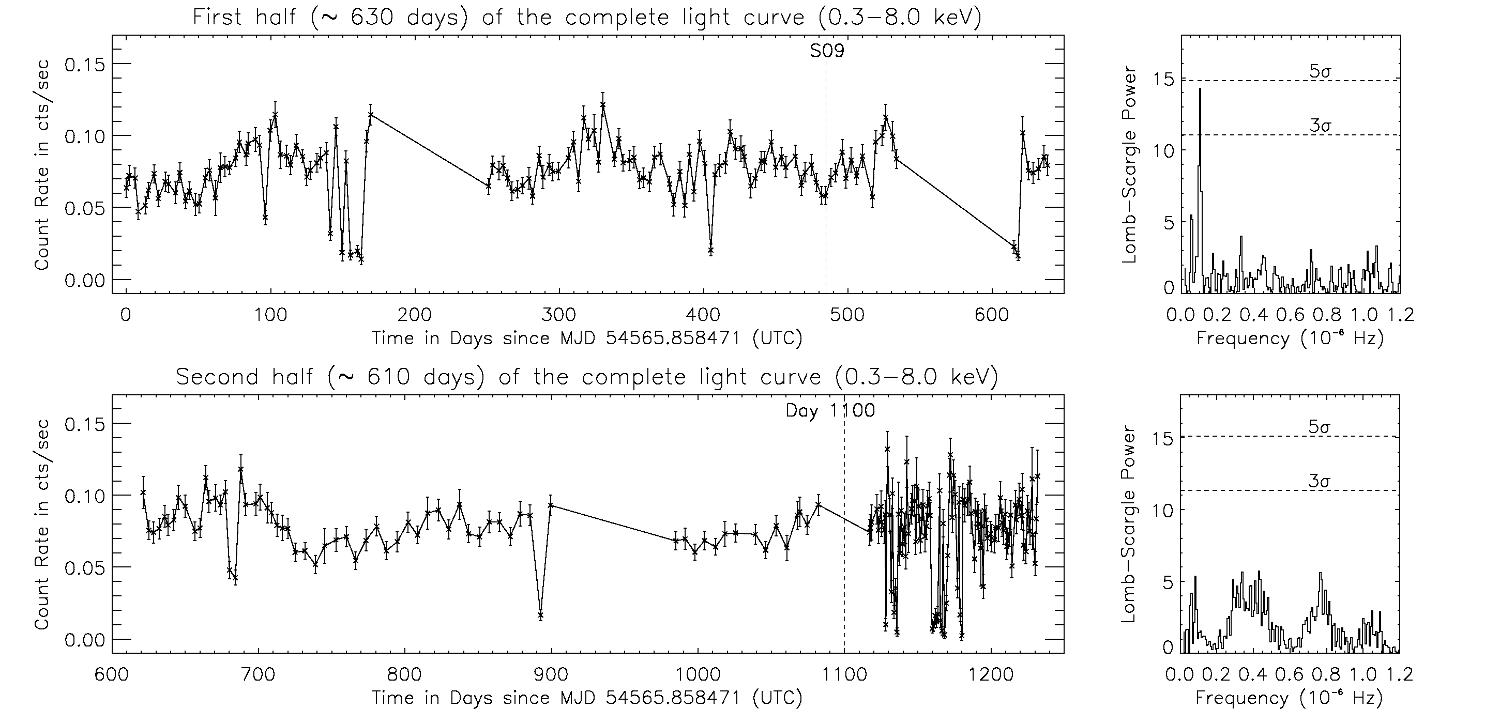}
\end{center}

{\textbf{Figure 2:} Top Left Panel: First half ($\approx$ 630
days) of the complete {\it Swift}/XRT X-ray light curve of the ULX NGC
5408 X-1 in the energy range 0.3-8.0 keV. Top Right Panel: The
Lomb-Scargle periodogram of the first half of the {\it Swift}/XRT
X-ray light curve of NGC 5408 X-1 (top left panel). The highest peak
corresponds to 112.6$\pm$4 days. Bottom Left Panel: Second half
($\approx$ 610 days) of the {\it Swift}/XRT X-ray light curve of the
ULX NGC 5408 X-1 in the energy range 0.3-8.0 keV. Note the erratic
dipping in the last continuous segment of the data. Bottom Right
Panel: The Lomb-Scargle periodogram of the second half of the
complete {\it Swift}/XRT X-ray light curve of NGC 5408 X-1 (bottom
left panel). Note the difference between the two periodograms: there
is no statistically significant period detected during the second half
of the light curve.}
\label{fig:figure2}
\end{figure*}


\section{Results: Timing Analysis}

This section is divided into four parts: (1) we show the complete
($\approx$ 1240 day temporal baseline) {\it Swift}/XRT X-ray (0.3-8.0
keV) light curve of NGC 5408 X-1, highlighting the most prominent
features, (2) we compute periodograms (Lomb-Scargle) of two different
portions of this complete light curve (one with strong
quasi-sinusoidal X-ray modulation and the other with apparently weaker
modulation), (3) we estimate the distribution of dips with orbital
phase and also construct an epoch folded light curve and finally, (4)
we study the energy dependence of the quasi-sinusoidal modulation and
the dips.

\subsection{X-ray Light Curve}

For each observation, we combined all the data to obtain an average
source and background count rate. The light curve of NGC 5408 X-1 in
the 0.3-8.0 keV energy range is shown in Figure 1. The background
count rate in the same energy range is also shown (red data points
with black error bars). The background is almost always negligible
compared to the source count rate. Roughly 40\% of these observations,
i.e., the first 113 observations of the present dataset, were analyzed
by S09. The vertical line marked as S09 indicates the end of the
observations used by S09 (see Figure 1).  S09 reported a 115.5 day
period in this data and concluded that it likely represents the
orbital period of the binary. The additional data provides new
insights about the system. A prominent feature of the full light curve
is the quasi-periodic recurrence of deep, sharp X-ray dips (though
with non-zero X-ray intensity during the dip minima). These dips are
highlighted with green circles in Figure 1. With the available
temporal baseline of $\approx$ 1240 days, we detected five epochs of
dips that recur roughly every 243 days. The approximate time intervals
between the dips are 251 days, 211 days, 276 days and 235 days.

\subsection{Timing Evidence for weakening of the 115 day modulation}


\begin{figure}[ht!]

\begin{center}
\includegraphics[width=3.5in, height=3.5in, angle=0]{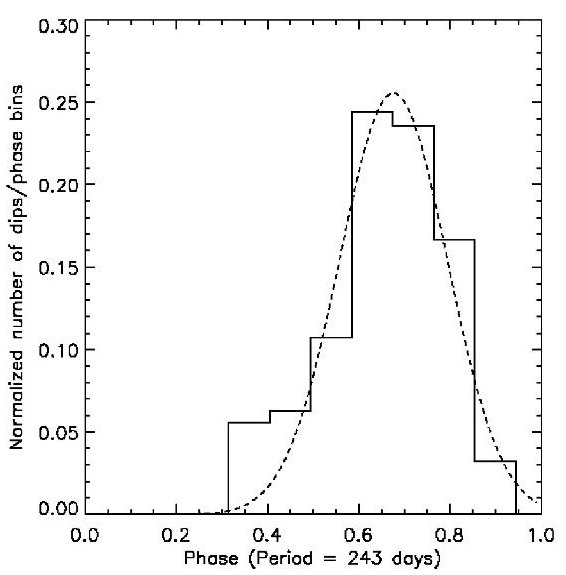}
\end{center}

{\textbf{Figure 3:} Frequency of the sharp dips as a function of the
orbital period (243 days). It is normalized by dividing each value by
the number of times each phase bin was observed. A total of 27 dips
were detected. 11 phase bins were used for the above distribution. The
best-fit Gaussian is also shown with a dashed line.}
\label{fig:figure3}
\end{figure}


In this section we present evidence for weakening of the 115.5 day
quasi-sinusoidal X-ray modulation during the second half ($\approx$
last 600 days) of the light curve shown in Figure 1.  We started our
analysis by dividing the light curve into roughly two equal
segments. We then directly compared the Lomb-Scargle periodograms
(Scargle 1982; Horne \& Baliunas 1986) of the two segments. We used
all the photons in the 0.3 - 8.0 keV energy range for this
analysis. The two periodograms with their respective light curves are
shown in Figure 2. The top left panel of the plot shows the light
curve of the first half of the complete light curve (first half of
Figure 1) and the top right panel shows the Lomb-Scargle periodogram
of this segment with confidence limits overlaid. The bottom left panel
shows the light curve of the second half of Figure 1 (note the x-axis)
and the bottom right panel shows the periodogram of the second half of
the complete light curve. The highest peak in the periodogram of the
first half of the light curve (top right panel of Figure 2)
corresponds to a period of 112.6$\pm$4 days. This is consistent with
the value reported by S09 (also see Han et al. 2012). Furthermore,
such a peak is not evident in the periodogram of the second half of
the data. We note that the reason for this difference is unlikely to
be purely statistical as the two portions of the light curve have
comparable signal-to-noise ratios, temporal baseline ($\approx$ 600
days each) and sampling rate (with the exception of the last
continuous segment of the light curve which has a higher sampling rate
of $\approx$ once per day). It is very likely that the drop in the
amplitude of the modulation in the second half of the light curve is
physical.


\begin{figure}[ht!]

\begin{center}
\hspace{-.25in}
\includegraphics[width=3.5in, height=3.in, angle=0]{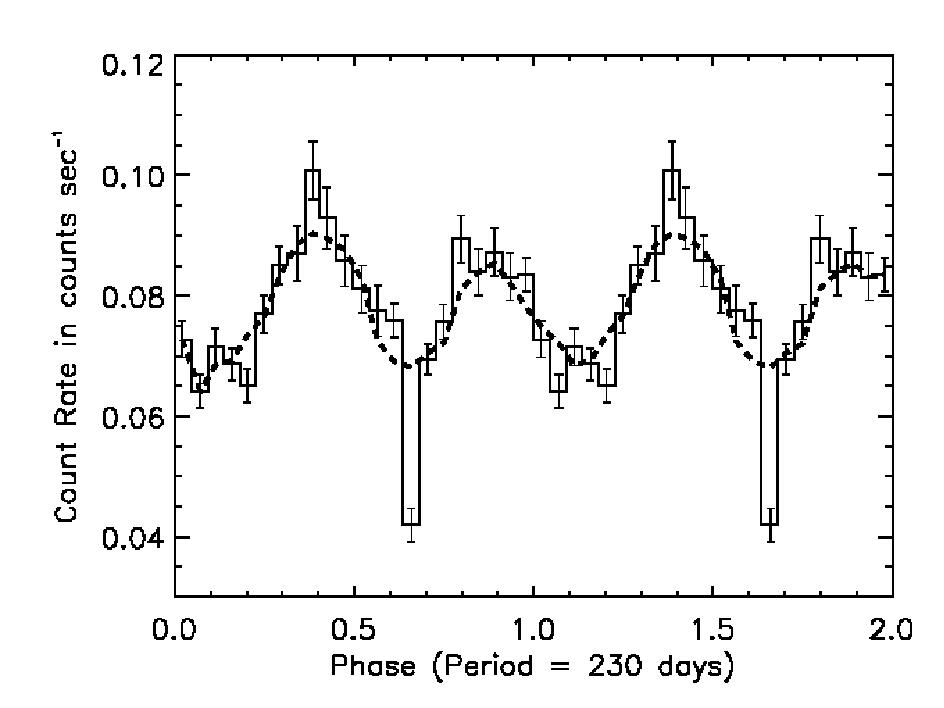}
\end{center}

{\textbf{Figure 4:} The epoch folded (period = 230 days) X-ray
(0.3-8.0 keV) light curve of NGC 5408 X-1. A total of 22 phase bins per
cycle were used and two cycles are shown for clarity. Also overplotted
is a running average curve to guide the eye. Each point is an average
of 5 neighboring bins. Both the smooth components (period of 115.5 days) and the sharp dips (quasi-period of 243$\pm$23 days) can be clearly seen.}
\label{fig:figure4}
\end{figure}

\subsection{Dip distribution and the folded light curve}
We have detected a total of 27 dips (see Figure 1). Each of these dips
was assigned a phase assuming a period of 243 days, the average time
interval between successive epochs of dipping. A normalized
distribution of the phases of these dips is shown in Figure 3. A
normalized distribution was obtained by dividing each value by the
number of times each phase bin was observed. This was done to
eliminate the bias of preferentially sampling certain phase
bins. Since the absolute orbital phase of the system is not known we
arbitrarily assign the zero phase to correspond to the start of the
monitoring observations. One can see that the dip phase distribution
is roughly Gaussian with a width (FWHM) of $\approx$ 0.24.

Another analysis tool to understand the nature of the modulations is
epoch folding of the light curve. In the case of NGC 5408 X-1, we have
detected five epochs of sharp dips. If these are associated with the
orbital period of the binary then the implied period is likely in the
range 243$\pm$23 days. The variance (23 days) on this value is rather
large, thus, the precise value of the orbital period to use for
folding the data is not known. Therefore, we used values between 220 -
270 days and then chose the value that resulted in the highest
modulation amplitude. This value is 230 days. Furthermore, it was
noted previously that the smooth, quasi-sinusoidal modulation
disappears/weakens in the latter half of the data. Therefore, we used
only the portion of the complete light curve that shows strong
evidence for both components, i.e., the first $\approx$ 700 days of
the data. The epoch folded light using all the photons in the 0.3-8.0
keV X-ray band is shown in Figure 4, and one can clearly see both
modulations.

\subsection{Energy dependence of the X-ray modulations}
\subsubsection{Energy-dependent quasi-sinusoidal modulation}
Using only part of the present dataset, S09 reported that the
amplitude of the quasi-sinusoidal X-ray modulation was energy
dependent, with the modulation amplitude decreasing with increasing
energy.  Here we attempt to systematically quantify the variability as
a function of energy. The procedure we carry out is as follows. We
obtain folded light curves in different energy bands. More
specifically, keeping the upper limit of the bandpass constant at 8.0
keV, we vary the lower limit of the bandpass from 0.3 to 2.0
keV. Again, since the 115.5 day modulation weakens during the later
half of the monitoring we used only the first $\approx$ 700 days of
data. A few folded profiles along with the best-fitting model (solid
line) are shown in Figure 5. Two cycles are shown for clarity. Here we
used only 13 phase bins per cycle as our goal is to model only the
overall quasi-sinusoidal modulation and not to study the subtle
features within the modulation profiles.

To each of these profiles we fit a model that includes two Fourier
components (the fundamental and the first harmonic), i.e., I = A +
B$\sin$2$\pi$($\phi$-$\phi_{0}$) +
C$\sin$4$\pi$($\phi$-$\phi_{1}$). All the fits give acceptable values
of reduced $\chi^{2}$ ($\approx$ 1 with 8 degrees of freedom). The
fractional amplitude for such a model is defined as f$_{amplitude}$ $=
( max(I) - min(I) ) / ( max(I) + min(I))$, and is an indicator of the
amount of variability in the source in the given energy range. Figure
6 shows the variation of the fractional amplitude (y-axis) as a
function of the lower limit of the bandpass considered
(x-axis). Clearly, the fractional amplitude of the X-ray modulation is
dependent on the energy range under consideration. The amplitude
decreases with increasing energy. These results are consistent with
those reported by S09.


\begin{figure*}

\begin{center}
\includegraphics[width=6.in, height=6.in, angle=0]{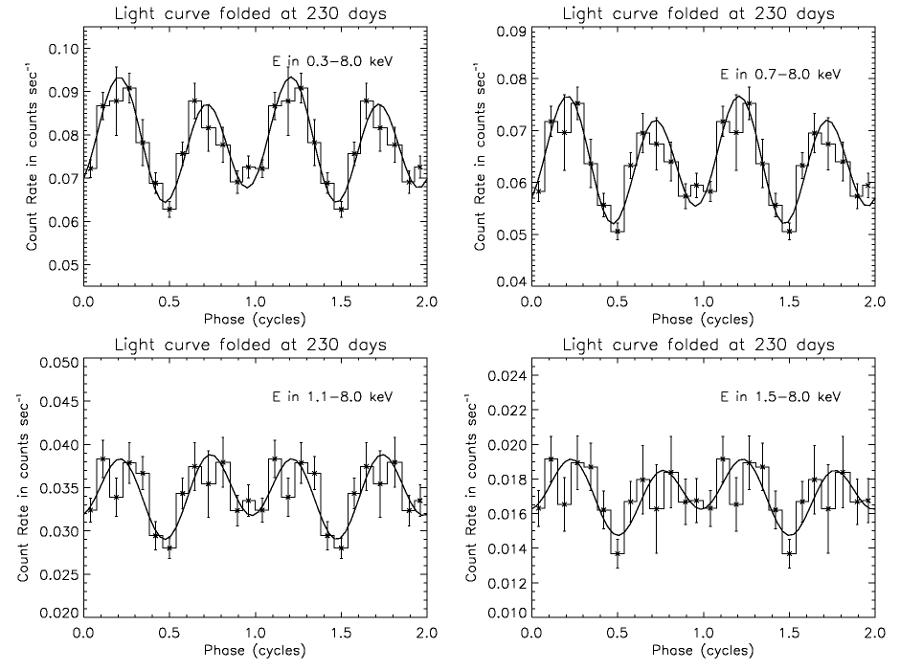}
\end{center}

{\textbf{Figure 5:} The epoch folded light curves of NGC 5408 X-1 in
varying energy bands. The light curves in the energy range 0.3-8.0
keV, 0.7-8.0 keV, 1.1-8.0 keV and 1.5-8.0 keV are shown in the top
left, top right, bottom left and the bottom right panels,
respectively. In each case, we used 13 phase bins per cycle and two
cycles are shown for clarity. To each of these profiles we fit a model
that includes two Fourier components (the fundamental and the first
harmonic), i.e., I = A + B$\sin$2$\pi$($\phi$-$\phi_{0}$) +
C$\sin$4$\pi$($\phi$-$\phi_{1}$). The best fitting curve is overlaid
(solid curve). Clearly, the modulation amplitude decreases with
increasing energy.}
\label{fig:figure5}
\end{figure*}

\subsubsection{Energy independent or dependent sharp dips? }

To study the energy dependence of the dips, we first extracted the
hardness ratio during these dips and elsewhere. We define this as the
ratio of the count rate in the hard band (1.0-8.0 keV) to the soft
band (0.3-1.0 keV). The top panel of Figure 7 shows the hardness ratio
as a function of time (in days). The middle and the bottom panels show
the light curves in the soft (0.3-1.0 keV) and the hard (1.0-8.0 keV)
bands, respectively. Owing to the lower count rates during the dips,
the error bars on the hardness ratio are large. It is not clear from
these plots alone whether the dips from NGC 5408 X-1 are energy
dependent. To explore further we obtained an average hardness ratio of
the dips before and after day 1100. These values are 1.18$\pm$0.17 and
0.95$\pm$0.19, respectively.  Within the error bars they are not only
consistent with each other, but are also consistent with the average
value of the non-dip observations. This suggests that the X-ray dips
are largely energy-independent.

\section{Results: Spectral analysis}

This section is divided into two parts: (1) we investigate the
differences in the spectra derived from two different portions of the
complete light curve, providing evidence for a change in the spectrum
of the source and (2) we compare the average X-ray spectrum of the
dips with the spectra derived from the rest of the observations.


\begin{table*}
    \caption{{ Summary of the best-fitting model parameters of the X-ray modulation profiles of NGC 5408 X-1}}\label{Table1} 
{\small
\begin{center}
    \begin{tabular}[t]{lcccccccc}
    \hline\hline \\
Bandpass (keV) & A\tablenotemark{1}(10$^{-2}$) & B\tablenotemark{1}(10$^{-2}$) & $\phi^{1}_{0}$ & C\tablenotemark{1}(10$^{-2}$) & $\phi^{1}_{1}$ &f\tablenotemark{2}$_{amplitude}$ & $\chi^2$/dof \\
	\\
    \hline \\
 0.3-8.0 & $7.82 \pm 0.09$ & $3.57 \pm 0.13$  & $0.88 \pm 0.05$ & $-1.20 \pm 0.13$ & $1.34 \pm 0.01$ & $0.184 \pm 0.021$ & 11.6/8 \\

	\\

 0.5-8.0 & $7.55 \pm 0.09$ & $3.76 \pm 0.14$  & $0.91 \pm 0.05$ & $-1.25 \pm 0.13$ & $0.84 \pm 0.01$ &$0.196 \pm 0.021$ & 13.4/8 \\

	\\

 0.7-8.0 & $6.40 \pm 0.09$ & $2.87 \pm 0.11$  & $0.86 \pm 0.07$ & $-1.03 \pm 0.12$ & $0.84 \pm 0.01$ & $0.191 \pm 0.023$ & 11.7/8 \\

	\\

 0.9-8.0 & $4.89 \pm 0.08$ & $1.71 \pm 0.08$ & $0.79 \pm 0.11$ & $0.68 \pm 0.11$ & $1.09 \pm 0.01$ & $0.163 \pm 0.025$ & 7.9/8 \\

	\\

 1.1-8.0 & $3.45 \pm 0.06$ & $1.40 \pm 0.07$ &  $0.70 \pm 0.11$ & $0.41 \pm 0.08$ & $0.60 \pm 0.01$ & $0.144 \pm 0.028$ & 12.7/8 \\

	\\

 1.3-8.0 & $2.39 \pm 0.05$ & $1.21 \pm 0.05$ & $0.71 \pm 0.10$ & $0.28 \pm 0.07$ & $0.60 \pm 0.02$ & $0.147 \pm 0.032$ & 9.8/8 \\
	
	\\

 1.5-8.0 & $1.71 \pm 0.04$ & $0.08 \pm 0.04$ & $0.81 \pm 0.12$ & $-0.16 \pm 0.06$ & $-0.13 \pm 0.03$ & $0.130 \pm 0.039$ & 8.5/8 \\

	\\

 1.7-8.0 & $1.29 \pm 0.04$ & $0.08 \pm 0.04$ & $0.81 \pm 0.11$ & $-0.12 \pm 0.05$ & $-1.13 \pm 0.03$ & $0.139 \pm 0.045$ & 6.3/8 \\

	\\

 2.0-8.0 & $0.90 \pm 0.03$ & $0.07 \pm 0.03$ & $0.78 \pm 0.10$ & $0.07 \pm 0.03$ & $0.61 \pm 0.04$ &$0.132 \pm 0.050$ & 6.9/8 \\

	\\

    \hline\hline
    \end{tabular}
\end{center}
}
\tablenotemark{1}{We fit the X-ray modulation profiles with a model consisting of two Fourier components. The mathematical form of the model is shown below: \\
\begin{center}
\begin{math} I = A + BSin2\pi(\phi-\phi_{0}) + CSin4\pi(\phi-\phi_{1}) \end{math}
\end{center}
where, A is the mean countrate while B and C are the amplitudes of the fundamental and the first harmonics, respectively. 
}\\
\tablenotemark{2}{This parameter is the fractional modulation amplitude and gives a quantitative measure of the amount of variation in the pulse profile. This is defined as follows: \\
\begin{center}
\begin{math} f_{amplitude} =  \frac {I_{max} - I_{min}} {I_{max} + I_{min}} \end{math}
\end{center}
}
\end{table*}


\subsection{Spectral evidence for a change in the absorption}

In this section we investigate the differences in the average spectra
of the source before and after day 1100. We chose this epoch because
of the apparent increase in the frequency of dips compared to the
earlier parts of the light curve.  For this purpose we used all but
the dip observations, i.e., the observations with count rate (0.3-8.0
keV) lower than 0.045 counts sec$^{-1}$ were excluded.  The number of
counts in any individual observation is too low (a few 10s of counts)
to extract a meaningful spectrum. This necessitated combining the
individual observations to obtain an average spectrum.  We did this
for all observations both preceding and following day 1100, resulting
in two spectra for comparison. We used the FTOOL {\it sumpha} to
combine the individual spectra. The two spectra in the X-ray energy
range of 0.5-8.0 keV are shown in Figure 8. The average spectrum prior
to day 1100 is shown in black while the data from after day 1100 is
shown in red. Both these spectra were binned to ensure a minimum of 50
counts in each spectral bin.

To quantify the spectra we fit them with a model that is often used to
describe the X-ray spectra of accreting black hole binaries: a
multi-colored disk and a power law, all modified by photoelectric
absorption. This same model has been used to fit the high-resolution
($\approx$ 100,000 counts) {\it XMM-Newton} spectra of NGC 5408 X-1
(Strohmayer et al. 2007; Dheeraj \& Strohmayer 2012). We used the
XSPEC (Arnaud 1996) spectral fitting package to fit all our
spectra. In terms of XSPEC models, we used {\it phabs*(diskpn+pow)}.
The X-ray spectrum of the dips (green points in Figure 8) suffers from
significant statistical uncertainty below 0.6 keV. Therefore, we used
the energy range of 0.6-8.0 keV for its spectral modeling. To be
consistent across all the X-ray spectra (as we will be comparing them
directly with each other) we chose the same energy range to model the
two average, non-dip spectra.  The results of the spectral fits are
summarized in Table 2.


\begin{table*}
  \caption{ Summary of the spectral modeling of NGC 5408 X-1. Best-fitting parameters for the {\it phabs$\ast$(diskpn+pow)} model are shown.}\label{Table2} \centering
{\small
  \begin{center}
    \begin{tabular}{lcccccccc}
    \hline\hline \\	
& & & & {\it phabs$\ast$(diskpn+pow)\tablenotemark{$\dagger$}} & & & \\
\\
    \hline\hline \\
  \hspace{7pt} Dataset & n$^{a}_{H}$ & N$^{b}_{disk}$($\times$10$^{-2}$)  & $\Gamma$\tablenotemark{c}  & N$^{d}_{pow}$($\times$10$^{-4}$) & Flux\tablenotemark{e}$_{0.6-8 keV}$  & Flux\tablenotemark{f}$_{Disk}$ & $\chi^2$/dof \\
	\\
    \hline \\
\hspace{8pt} Average & & & & & & & \\
\hspace{6pt} Spectrum         & 0.30$^{+0.02}_{-0.02}$& 1.57$^{+0.22}_{-0.27}$ & 2.96$^{+0.07}_{-0.07}$  & 7.69$^{+0.49}_{-0.48}$  & 3.68$^{+0.21}_{-0.23}$$\times$10$^{-12}$ & 1.69$^{+0.29}_{-0.33}$$\times$10$^{-12}$ & 233/179 \\
\hspace{-3pt} Before day 1100 & & & & & & & \\
\\
\hspace{8pt} Average  & & & & & & & \\
\hspace{6pt} Spectrum	& 0.44$^{+0.04}_{-0.04}$& 4.27$^{+0.83}_{-0.86}$ & 2.97$^{+0.17}_{-0.15}$  & 8.64$^{+1.42}_{-1.19}$  & 6.71$^{+0.70}_{-0.77}$$\times$10$^{-12}$ & 4.35$^{+0.89}_{-0.77}$$\times$10$^{-12}$ & 112/87 \\ 
\hspace{-3pt} After day 1100  & & & & & & & \\
\\
\hspace{8pt} Average  & & & & & & & \\
\hspace{18pt} Dip	& 0.31$^{+0.15}_{-0.18}$& 0.45$^{+0.54}_{-0.42}$ & 2.92$^{+0.61}_{-0.53}$  & 1.78$^{+1.05}_{-0.70}$  & 0.96$^{+0.62}_{-0.48}$$\times$10$^{-12}$ & 0.51$^{+0.82}_{-0.82}$$\times$10$^{-12}$ & 17/17  \\ 
\hspace{8pt} Spectrum & & & & & & & \\
\\
    \hline\hline 
    \end{tabular}
\end{center}
\begin{flushleft}{
\tablenotemark{a}{Total column density of hydrogen along the line of sight including the Galactic extinction (in units of 10$^{22}$cm$^{-2}$). We used the {\it phabs} model in XSPEC.}
\tablenotemark{b}{The normalization of the disk component. We used the {\it diskpn} model in XSPEC. The inner radius of the disk was fixed at 6GM/c$^{2}$. The disk temperature was fixed at 0.149 keV to avoid parameter degeneracy (see text).}
\tablenotemark{c}{The photon index of the power law.}
\tablenotemark{d}{The normalization of the power law component. We used {\it pow} model in XSPEC.}
\tablenotemark{e}{The total unabsorbed X-ray flux (in units of ergs cm$^{-2}$ s$^{-1}$) in the energy range of 0.6-8.0 keV.}
\tablenotemark{f}{The disk contribution to the total X-ray flux (in units of ergs cm$^{-2}$ s$^{-1}$) in the energy range of 0.6-8.0 keV.}
}\end{flushleft}
}
\end{table*}


Even without detailed modeling it can be seen straightaway that there
are significant differences between the two spectral below 1.5 keV
(black and red symbols in Figure 8). A closer look at the unfolded
spectra reveals that the disk black body component dominates at
energies below 2 keV while the powerlaw component dictates the nature
of the spectrum at higher energies; and the photoelectric absorption
is stronger at lower energies.  Therefore, the apparent differences
between the two spectra can either be due to a change in the
absorption column density ({\it phabs}) or to changes in the disk
properties, viz., the disk temperature, and/or the disk
normalization. The quality of the data does not allow us to
independently constrain each model's parameters. To break this
degeneracy, either one of the {\it diskpn} model parameters (the disk
temperature or the disk normalization) or the column density had to be
frozen. A careful analysis of the spectra reveals stronger statistical
evidence for a varying column density than varying disk properties.

To illustrate this we simultaneously, but independently, fit the model
to both the spectra (without freezing any model parameters except for
the inner radius of the accretion disk). We then obtain confidence
contours ($\chi^2$) between the column densities, the disk
temperatures and the disk normalizations of the spectra before and
after day 1100. These confidence contours are shown in Figure 9. The
top panel shows the confidence contours between the column density of
hydrogen prior to day 1100 (X-axis) and after day 1100
(Y-axis). Similarly, the bottom left and the bottom right panels show
the contours between the disk temperatures and the disk
normalizations, respectively. In all the cases, the black, the red and
the green represent the 1$\sigma$, 2$\sigma$ and the 3$\sigma$
confidence contours, respectively. In each case, the diagonal line
shows the locus of points where the value on the X-axis equals the
value on the Y-axis.


\begin{figure}[ht!]

\begin{center}
\hspace{-.25in}
\includegraphics[width=3.5in, height=3.75in, angle=0]{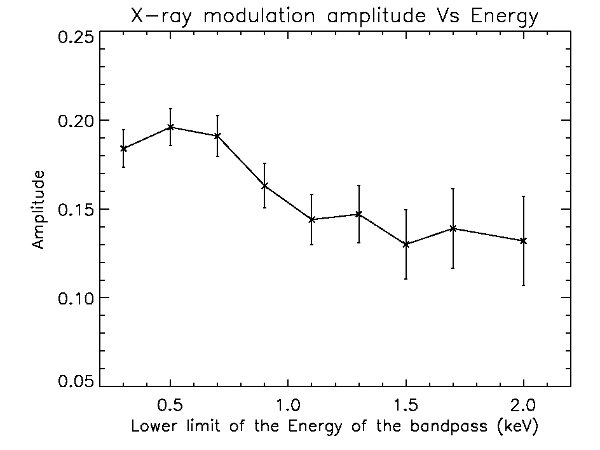}
\end{center}

{\textbf{Figure 6:} The fractional amplitude of the smooth component of the X-ray modulation
as a function of the lower limit of the band pass is shown. The upper
limit of the bandpass was fixed at 8 keV. The fractional amplitude of
the mathematical form used to model the epoch folded light curves of
NGC 5408 X-1 (Figure 4) is defined as f$_{amplitude}$ $= ( max(I) -
min(I) ) / ( max(I) + min(I))$, where I is the X-ray count rate.  }
\label{fig:figure6}
\end{figure}


The amount of deviation of the confidence contours from this line
indicates the significance of the variation of a given parameter
between the two epochs. In other words, the larger the deviation the
stronger is the evidence for a change in the given parameter. Clearly,
the statistical evidence for an increase in the column density after
day 1100 is very strong (top panel: strong deviation from the straight
line) compared to that for changes in the disk temperature (bottom
left panel) and the disk normalization (bottom right panel). In
addition, the source has been observed on multiple occasions with {\it
XMM-Newton} over the past ten years. The disk temperature on those
occasions has remained roughly constant (Soria et al. 2004; Strohmayer
\& Mushotzky 2009; Dheeraj \& Strohmayer 2012). Furthermore, in
accreting galactic black hole binaries, changes in the accretion disk
temperature are often accompanied by significant variations in the
X-ray light curve (McClintock \& Remillard 2006). Clearly, this is not
seen in the light curve of NGC 5408 X-1. These arguments suggest that
the disk temperature has likely remained constant. Therefore, for
subsequent analysis we assumed a constant disk temperature of 0.149
keV, the average value reported from the analysis of the high-quality
{\it XMM-Newton} data (Dheeraj \& Strohmayer 2012). The model then
gives a reasonable fit in both cases (before and after day 1100) with
reduced $\chi^{2}$ of 1.3 (179 degrees of freedom) and 1.3 (87 degrees
of freedom). The best-fitting model parameters for the two average
spectra are shown in the first and the second row of Table 2,
respectively.


\begin{figure*}

\begin{center}
\includegraphics[width=6.15in, height=6.65in, angle=0]{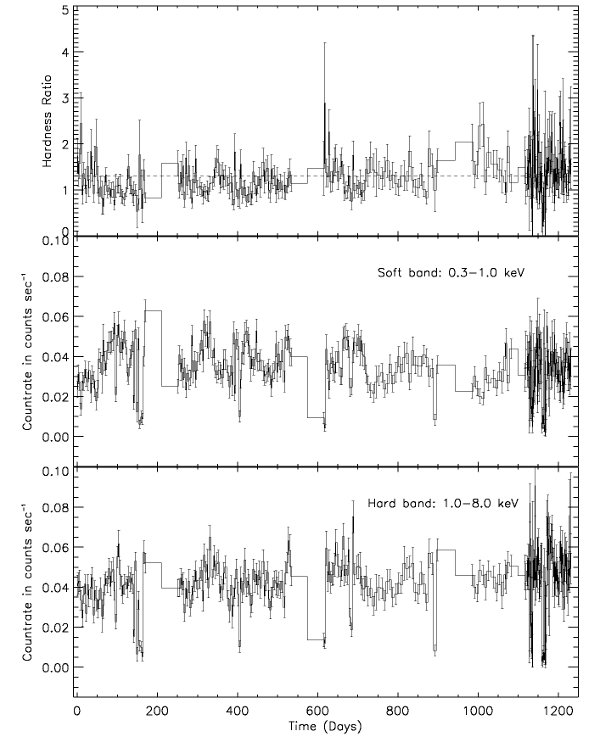}
\end{center}

{\textbf{Figure 7:} Top panel: Time history of the hardness
ratio (ratio of the count rate in the hard band (1.0-8.0 keV) to the
soft band (0.3-1.0 keV)). The mean value is indicated by the dashed
line. Middle panel: The light curve of the source in the soft
band, i.e., 0.3-1.0 keV. Bottom panel: The light curve of the
source in the hard band, i.e., 1.0-8.0 keV.}
\label{fig:figure7}
\end{figure*}


The most prominent difference between the two spectra is the amount of
photoelectric absorption. The total hydrogen column density before and
after day 1100 was $0.30^{+0.02}_{-0.02}$$\times$10$^{22}$cm$^{-2}$
and $0.44^{+0.04}_{-0.04}$$\times$10$^{22}$cm$^{-2}$,
respectively. That is, the latter X-ray spectrum is more absorbed than
the former. The estimated total unabsorbed flux in the 0.6-8.0 keV
range before and after day 1100 is
$3.7^{+0.2}_{-0.2}$$\times$10$^{-12}$ergs cm$^{-2}$ s$^{-1}$ and
$6.7^{+0.7}_{-0.8}$$\times$10$^{-12}$ergs cm$^{-2}$ s$^{-1}$,
respectively. This suggests a factor of 2 increase in the unabsorbed
X-ray flux during the last continuous segment of the light
curve. Another important thing to note is the fraction of the disk
contribution to the total flux before and after day 1100. The values
are $1.7^{+0.3}_{-0.3}$$\times$10$^{-12}$ergs cm$^{-2}$ s$^{-1}$ and
$4.6^{+0.9}_{-0.8}$$\times$10$^{-12}$ergs cm$^{-2}$ s$^{-1}$,
respectively. This implies that the disk fraction of the total X-ray
flux (0.6-8.0 keV) before and after day 1100 is 46$\pm$9\% and
68$\pm$15\%, respectively. In other words, the latter spectrum appears
more disk-dominated.

\subsection{Spectra during the sharp dips versus elsewhere}
The dips detected here occur both prior to and after day 1100 (see
Figure 1); and the individual dip observations have too few counts (a
few 10s) to extract meaningful spectra. However, it was noted earlier
that the average hardness ratio of both the dips before and after day
1100 are comparable, suggesting a plausible common physical
origin. Therefore, we combined all the dip observations (both prior to
and after day 1100) and obtained an average dip spectrum. This is
shown in Figure 8 (green data points) along with the two other average
spectra. The spectrum was binned to ensure a minimum of 25 counts in
each spectral bin.

We began by fitting the dip spectrum with the same model ({\it
phabs*(diskpn+pow)}) used to fit the two spectra described in the
previous section. This model gives a good fit with a reduced $\chi^2$
of 1.0 (17 degrees of freedom). The best-fit parameters are shown in
the last row of Table 2. We find that the best-fit parameters of the
dip spectrum are consistent with the values for the two average
non-dip spectra derived earlier. This is also consistent with the
simpler hardness ratio analysis, i.e., the dips are consistent with
being energy independent.


\begin{figure}[ht!]

\begin{center}
\includegraphics[width=3.5in, height=3.15in, angle=0]{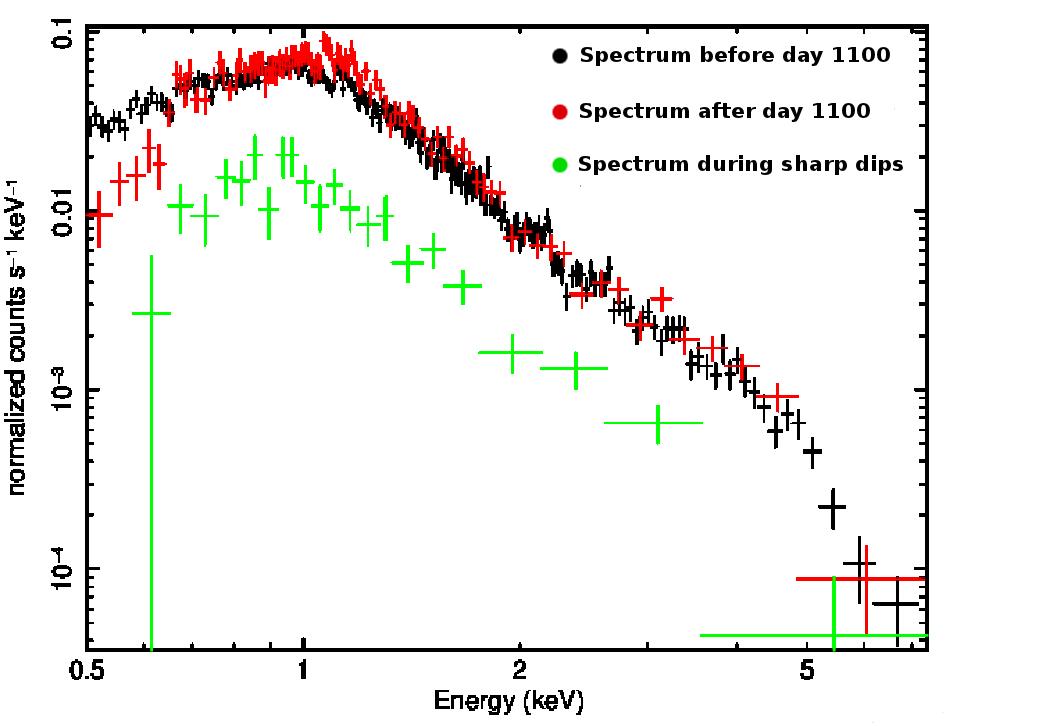}
\end{center}

{\textbf{Figure 8:} The {\it Swift}/XRT spectra of NGC 5408 X-1 in the
energy range 0.5-8.0 keV. The {\it black} points show the average
spectrum prior to day 1100 (as marked by the vertical line in Figure
1). Shown in {\it red} is the average spectrum of the source after day
1100 epoch. There are clear differences between the two spectra,
especially at lower energies (below 1.5 keV). The {\it green} data
points denote the average spectrum during the sharp dips. The two
spectra, i.e., the spectra before and after day 1100, are binned to
ensure a minimum of 50 counts per spectral bin, while the dip-spectrum
is binned to ensure a minimum of 25 counts per spectral bin. }
\label{fig:figure8}
\end{figure}


The dip spectra of X-ray binaries are often characterized using more
complex models involving either an absorbed-plus-unabsorbed
(hereafter, A+U) approach (e.g., Parmar et al. 1986, Courvoisier et
al. 1986) or a progressive covering approach (e.g., Church et
al. 1997, 1998, 2001).  To explore the A+U approach we used the
following model in {\it XSPEC}, {\it phabs1*const1*(diskpn+powerlaw) +
phabs2*const2*(diskpn+powerlaw)}, where the first and second terms
represent the unabsorbed and absorbed components, respectively. In
this case the only free parameters are the two normalization constants
({\it const1} and {\it const2}) and the column density associated with
{\it phabs2}. The parameters describing the shape of the spectrum are
all fixed at the average best-fit values derived from the two non-dip
spectra (first and second rows of Table 2).  This model gives an
acceptable fit with a reduced $\chi^2$ of 0.9 with 19 degrees of
freedom.  The best-fit values of the normalization of the unabsorbed
component, normalization of the absorbed component and the column
density of the absorbed component are 0.2, 4.1$\times$10$^{-4}$ and
2.7$\times$10$^{22}$cm$^{-2}$, respectively.  A simple interpretation
in the context of this model is that during dips the direct X-ray
emission is completely obscured/absorbed, and what remains is largely
scattered (in an energy independent fashion) into our line of sight
from more extended regions around the source.

Partial covering models have been quite successful in describing the
spectral evolution with intensity during dipping (see for example
Church et al. 1998).  In this approach the spectral evolution as
dipping becomes stronger is accounted for by the progressive covering
of, typically, one of the spectral components.  The implication is
that the progressively covered component is spatially extended
compared to the other component.  A perceived strength in favor of
these models is that scattering--the details of which are not often
specified in the A+U approach--is not required. In the case of LMXB
neutron star dippers it is the coronal (power-law) component that is
progressively covered, whereas in some black hole systems successful
modeling requires partial covering of the thermal (disk black body)
component (Church 2001).  We found that the dip spectrum of NGC 5408
X-1 could not be well described by progressive covering of either
spectral component by itself, but that progressive covering of the sum
of both components provided as good a fit as the A+U modeling.  The
reason for this is two-fold, 1) the dip spectrum is consistent with
being energy independent, that is, it has the same shape as the
non-dip spectra, and 2) both spectral components contribute a
comparable amount to the total flux.  Thus, it is very difficult to
partially cover only one component and still maintain the same
spectral shape.  Partial covering of just the power-law component does
a better job than covering of only the {\it diskpn} component, but in
each case the fits are not as good as the A+U case. Covering of the
sum of the components provides a good fit. The implication in this
case is that the emission components are extended but co-spatial. The
dipping then corresponds to complete obscuration of about 4/5 of the
extended source (in an average sense over all dips).

The conclusion from either modeling framework is that an absorbed
component is not seen directly, only an unabsorbed component.
Finally, we note that due to the limited statistical quality of the
dip spectrum and the fact that only a single dip spectrum could be
meaningfully extracted, it is difficult to be more precise about the
exact physical nature of the dipping.


\begin{figure*}

\begin{center}
\includegraphics[width=6.in, height=6.0in, angle=0]{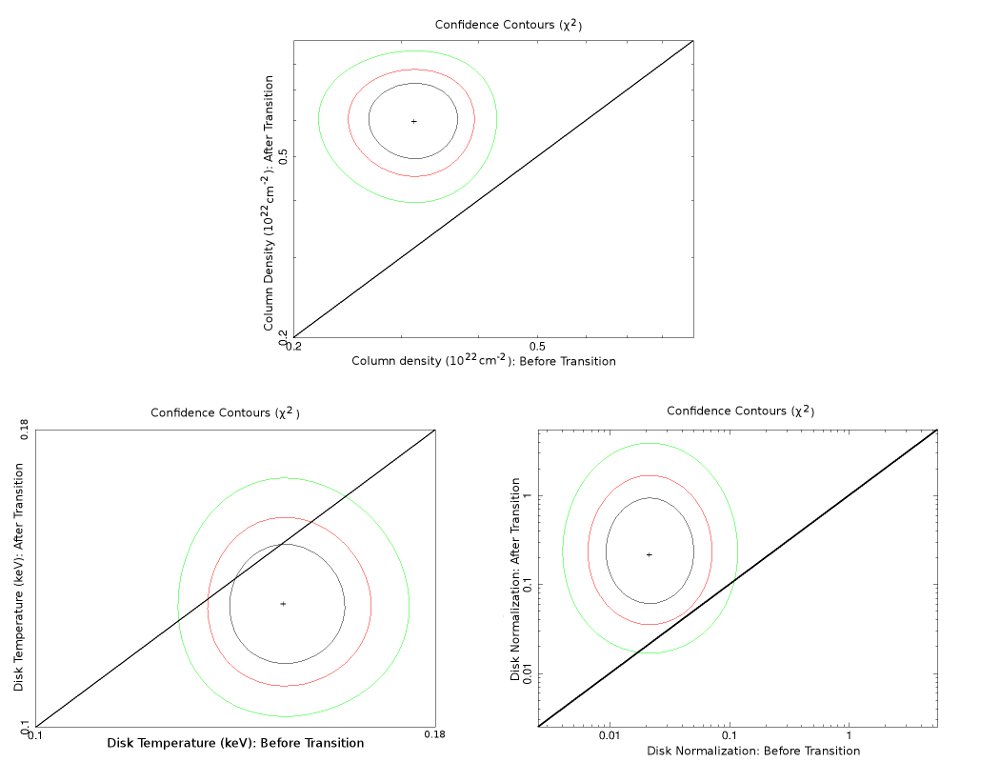}
\end{center}

{\textbf{Figure 9:} Top Panel: The confidence contours of the
total hydrogen column densities along the line of sight (n$_{H}$)
before (X-axis) and after (Y-axis) day 1100. Bottom Left Panel:
The confidence contours of the disk temperatures before (X-axis) and
after (Y-axis) day 1100. Bottom Right Panel: The confidence
contours of the disk normalizations before (X-axis) and after (Y-axis)
day 1100. The black, red and the green curves in each panel represent
the 1$\sigma$, 2$\sigma$ and the 3$\sigma$ confidence
contours. Contours were obtained from fits to the non-dip spectra in
both cases. The straight line in each case represent the locus of all
points with X = Y. For a given parameter, if the contours do not
overlap with this line then that is a strong indication for a change
in the parameter value.}
\label{fig:figure9}
\end{figure*}


\section{Discussion}
In this article we have studied the long-term X-ray variability of the
ULX NGC 5408 X-1 using 1240 days of {\it Swift}/XRT monitoring
data. In addition to the periodic (115.5 days), smooth modulation
reported by S09, the source also exhibits quasi-periodic sharp dips
that occur on average every 243$\pm$23 days, roughly twice the 115.5
day period. As discussed in \S 1, such orbit-phase dependent X-ray
dipping has been well-studied in the case of accreting X-ray binaries
and is usually ascribed to periodic obscuration of the X-ray source by
an intervening medium, viz., a bulge at the edge of the accretion
disk, the wind from the companion star or the accretion stream. In
those cases, dipping has been shown to be an excellent tracer of the
orbital motion of the binary. Given that NGC 5408 X-1 is very likely
an accreting black hole system, it is conceivable that a similar
phenomenon causes the observed dipping. If that were indeed the case,
then the detected period of 243$\pm$23 days likely represents the
orbital motion of the X-ray binary.

NGC 5408 X-1 is the only known ULX system that exhibits two types of
dipping behavior, smooth \& sharp, simultaneously. The basic
properties of the smooth component are: 1) it occupies a significant
portion ($\approx$50\%) of the putative orbital cycle of 243$\pm$23
days (see Figure 4), 2) it is energy-dependent with the modulation
amplitude decreasing with increasing energy (see Figures 5 \& 6) and
3) the modulation amplitude is variable, i.e., the smooth component
modulation is weaker during the second half of the {\it Swift}
monitoring data (see Figure 2). On the other hand, the sharp X-ray
dips detected from NGC 5408 X-1 are not strictly periodic and are
distributed over an orbital phase of $\approx$ 0.24 (see Figure 3). In
addition, they are consistent with being energy-independent (see
Figure 8 \& Table 2). We now discuss these observed properties within
the framework of an accreting X-ray binary scenario.

\subsection{Roche Lobe Accretion: Eccentric accretion disk and
stream-disk interaction} 

Assuming that the recurrence period of the X-ray dips is close to the
orbital period of a Roche-lobe filling binary, the density of the
companion star in NGC 5408 X-1 can be constrained using the formula,
$\rho$ $\simeq$ 0.2(P$_{days}$)$^{-2}$ g cm$^{-3}$ (Frank et
al. 2002). For a period (P$_{days}$) of 243 days, this would imply
that the companion star has a mean density of 3.4$\times$10$^{-6}$ g
cm$^{-3}$. This value is consistent with a recent study by Gris{\'e}
et al. (2012), who extracted the UV/optical/NIR spectral energy
distribution (SED) of the optical counterpart of NGC 5408 X-1. They
find that the SED is consistent with a massive B0I super-giant
star. Numerical hydrodynamical simulations of stream-fed accretion in
low mass ratio binaries show that the accretion disk can be tidally
distorted (Armitage \& Livio 1996). This leads to the formation of a
secondary bulge in the disk in addition to the primary bulge at the
stream-disk impact site. In other words, tidal effects can lead to the
formation of two broad bulges above the disk mid-plane along the outer
rim of the accretion disk, separated by an orbital phase of $\approx$
0.5 (see Figure 5 of Armitage \& Livio 1996). This process may explain
the observed modulations in NGC 5408 X-1. The modulation profile of the smooth component, including the phase
separation between the two peaks, is in accord with the predicted
X-ray variations from high inclination X-ray binaries (see the right
panels of Figure 6 of Armitage \& Livio 1996). Similar double-peaked
modulation profiles have also been observed from neutron star LMXBs
including X 1916-053 (Smale et al. 1988; Homer et al. 2001), X
0748-676 (Parmar et al. 1988) and 4U 1822-37 (White \& Holt
1982). Again, in those cases the presence of a secondary bulge,
presumably due to the tidal effects of the companion, has been
suggested.  In this context the nature of the accretor (neutron star
or black hole) is largely irrelevant, it is simply the tidal effects
on the disk that are important in producing a secondary bulge.


\begin{figure*}

\begin{center}
\includegraphics[width=4.75in, height=5.95in, angle=0]{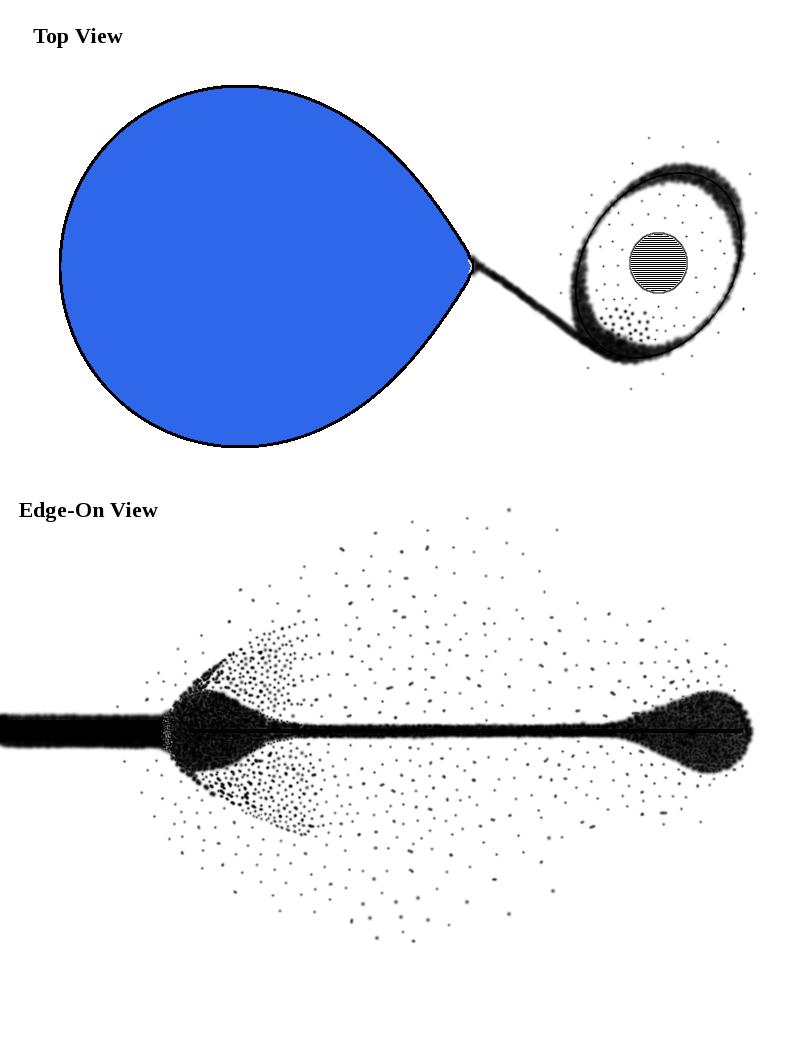}
\end{center}

{\textbf{Figure 10:} An idealized depiction of a possible accretion
geometry for NGC 5408 X-1. {\it Top Panel:} A face-on (top) view of
the binary system. The optical star (possibly a B0I supergiant,
Gris{\'e} et al. 2012) is shown in blue (left), while the eccentric
accretion disk with two obscuring regions is shown on the right hand
side. The X-ray source is indicated by a circular region at the center
of the disk. {\it Bottom Panel:} An edge-on view of the accretion
disk. The accretion stream from the companion approaches from the left
hand side. Again the two bulges are shown. The ``clumps'', produced
due to the splashing of the accretion stream at the edge of the disk,
are also shown (scattered black points). See also Figure 5 of Armitage
\& Livio (1996) and the bottom panel of Figure 7 of Armitage \& Livio
(1998).}
\label{fig:figure10}
\end{figure*}


Further simulations by Armitage \& Livio (1998) show that the fate of
the material in the accretion stream after it impacts the disk depends
on the cooling efficiency within the shock-heated gas created by the
impact. If the cooling is efficient, which is generally the case for
low mass accretion rates ($\la$ 10$^{-9}$ M$_{\odot}$ yr$^{-1}$) (see
Armitage \& Livio 1998 for details), the material continues to freely
overflow above and below the accretion disk. This ballistic stream
will reimpact the disk near the point of its closest approach to the
compact source where it forms an obscuring bulge (See also Frank et
al. 1987; Lubow 1989). However if the cooling is inefficient which is
thought to be the case for high mass accretion rates, the stream-disk
interaction leads to a flow that is better described as an
``explosion'' at the point of the impact, and leads to a more
vertically extended distribution of scattered material (clumps: See
the bottom panel of Figure 7 of Armitage \& Livio 1998). Such clumps
could account for the sharp dips observed in NGC 5408 X-1.  Given its
high luminosity and hence high inferred accretion rate (average X-ray
luminosity of $\approx$ 10$^{40}$ ergs s$^{-1}$) it is plausible that
the cooling at the stream-disk impact site is inefficient. Therefore,
one would expect a distribution of splashed material (clumps) around
the primary bulge. This is in reasonable agreement with the modulation
profile of NGC 5408 X-1 where the sharp dips occur predominantly at
one of the minima which is consistent with the phase of the
stream-disk impact site (primary bulge).

In summary, the observed modulation profile of NGC 5408 X-1 appears
consistent with the predicted variations from Roche-lobe accreting
binaries with low mass ratios, $q = M_{donor} / M_{accretor}$
(Armitage \& Livio 1996, 1998). The smooth component is produced by
absorption/obscuration by the two spatially distinct bulges along the
outer rim of the accretion disk while the sharp dips are possibly
produced due to absorption by clumps of material produced by the
stream-disk impact.  A schematic depiction of a possible source
geometry for NGC 5408 X-1 based on this scenario is shown in Figure
10. The top panel shows the distribution of material in the binary
viewed from above (face on) while the bottom panel shows the accretion
disk viewed edge-on.

The sharp dips seen in NGC 5408 X-1 appear similar to those observed
in the black hole LMXB GRO J1655-40 (Kuulkers et al. 2000), however,
GRO J1655-40 does not show a smooth periodic modulation as in NGC 5408
X-1. This may be due to the inclination of GRO J1655-40 ($i$ =
70.2$\pm$2 $^{\circ}$: Greene et al. 2001), as simulations show that
the bulge material has a limited extent above the disk.  However,
stream material splashing on to the edge of the accretion disk is able
to reach higher above the plane of the accretion disk (higher than the
bulges) so that in principle sharp dips can still occur away from the
plane of the accretion disk (Amritage \& Livio 1998). While the
morphology of the sharp X-ray dips in NGC 5408 X-1 somewhat resembles
partial eclipse profiles, we note that this is unlikely to be the case
due to the fact that the sharp dips are not strictly periodic. That
is, the large variance on the dip period (see Figure 3) strongly
indicates that we are not seeing eclipses.

If these two arguments are correct then a rough constraint on the
inclination angle can be derived, as the inclination angle has to be
less than some upper limit so that eclipses are not seen but must be
large enough so that the absorbing/obscuring effects of the disk
bulges are not completely eliminated. Assuming the donor star fills
its Roche lobe, an upper limit on the inclination angle (such that
eclipses are not seen) can be derived that depends on the orbital
period and the mass ratio of the binary (see Eggleton 1983; Rappaport
\& Joss 1986 for the appropriate formulae). For an orbital period of
243 days with a B0I donor of $\approx$ 10M$_{\odot}$, the upper limits
on the inclination are 75$^{\circ}$ and 85$^{\circ}$ assuming the
compact source is a massive stellar-mass black hole (mass $\approx$
50M$_{\odot}$: Middleton et al. 2011) or an intermediate-mass black
hole (mass $\approx$ 1000M$_{\odot}$: Strohmayer et al. 2009),
respectively. If we further assume that the absence of a smooth
modulation in GRO J1655-40 provides a rough lower limit of
$70^{\circ}$, we obtain an inclination angle for NGC 5408 X-1 in the
range from 70 - 85$^{\circ}$.  While this is based on a number of
assumptions we note that it is broadly consistent with inclination
angles inferred for other dipping binaries (see Ritter \& Kolb 2003).

Finally, we note that while the sharp dips seen in NGC 5408 X-1 are
similar to the Type-B dips observed in Cygnus X-1 (see \S 1 for
details on the sharp dips from Cygnus X-1), the high X-ray luminosity
and thus high inferred mass accretion rate, is much easier to
accommodate via Roche overflow than wind-fed accretion.

\subsection{Alternative Scenarios}

Alternatively, it could be that the observed X-ray modulations from
NGC 5408 X-1 are not due to the orbital motion of the X-ray source in
a binary (Foster et al. 2010). X-ray modulations on ``super-orbital''
periods (periods longer than the binary orbital period) of the order
of a few days to a few hundred days have been reported from accreting
X-ray binaries (e.g., Wen et al. 2006; Sood et al. 2007). In a
sub-sample of these sources consisting of Her X-1 (P$_{Superorbit}$ =
35 days), SS 433 (P$_{Superorbit}$ = 164 days), LMC X-4
(P$_{Superorbit}$ = 30.4 days) and SMC X-1 (P$_{Superorbit}$ $\approx$
55 days), it is likely that the observed X-ray modulation is due to
the periodic obscuration of the central X-ray source by a tilted,
precessing accretion disk. The basic idea here is that the accretion
disk is tilted or warped with respect to the orbital plane of the
binary. The tidal effects of the companion star then force the
accretion disk to precess about an axis normal to the binary orbit. As
the tilted disk precesses, the effective area of the disk obscuring
the central X-ray source can vary, producing modulations at or near
the precession period. Furthermore, the radiation force from the X-ray
source can warp the outer regions of the accretion disk (Ogilvie \&
Dubus 2001; Ogilvie 2002) leading to two spatially distinct density
enhancements. This might possibly account for the twin-peaked X-ray
modulation curve of NGC 5408 X-1. In summary, the theory of tilted,
warped accretion disks may be adaptable to the present scenario to
perhaps explain the observed properties.

\section{Summary}

We have presented results from long-term X-ray (0.3-8.0 keV)
monitoring of the ULX NGC 5408 X-1 using {\it Swift}/XRT. Our primary
results are: (1) the discovery of sharp, energy-independent dips (a
total of 5 dip epochs) in the X-ray intensity that recur roughly every
243 days and (2) the detection of a smooth, quasi-sinusoidal
modulation of the X-rays which appears to weaken during the second
half of the monitoring program. We interpret these findings in the
context of orbital motion in a Roch-lobe overflow binary with a period
comparable to the dip-recurrence period (243$\pm$23 days; see Figure
10), however, it is also possible that a precessing accretion disk (a
``super-orbital'' phenomenon) can cause a similar X-ray
modulation. Further X-ray monitoring of NGC 5408 X-1 with {\it Swift}
is warranted to more firmly establish that the X-ray dips are indeed
associated with the orbital period of the binary.


\vfill\eject

\end{document}